\documentclass[floatfix,aps,pra,twocolumn,showpacs,amsmath,preprintnumbers,nofootinbib,superscriptaddress]{revtex4} 

\usepackage{epsfig,verbatim}
\usepackage{amssymb}
\usepackage{graphicx}
\usepackage{color}

\def\scn#1#2{\section{#1}\lb{#2}}
\def\sscn#1#2{\subsection{#1}\lb{#2}}
\def\bfl{\begin{flushleft}}
\def\efl{\end{flushleft}}
\def\bfr{\begin{flushright}}
\def\efr{\end{flushright}}
\def\bc{\begin{center}}
\def\ec{\end{center}}
\def\be{\begin{equation}}
\def\ee{\end{equation}}
\def\ba{\begin{eqnarray}}
\def\ea{\end{eqnarray}}
\def\baa#1{\begin{array}{#1}}
\def\eaa{\end{array}}
\def\bw{\begin{widetext}}
\def\ew{\end{widetext}}
\def\nn{\nonumber }
\def\lb#1{\label{#1}}
\def\bit{\begin{itemize}}
\def\eit{\end{itemize}}
\def\bco{}

\def\schrod{Schr\"odinger  }

\def\acorrx{{\cal C}_{\sigma_x \sigma_x}}
\def\acorrz{{\cal C}_{\sigma_z \sigma_z}}
\def\corrxz{{\cal C}_{\sigma_z \sigma_x}}
\def\corryz{{\cal C}_{\sigma_z \sigma_y}}
\def\densnnorm{\hat{\Omega}}
\def\rabi{\Delta}

\begin{document}

\preprint{\small Phys. Rev. A 91 (2015) 062108 [arXiv:1412.5782]}

\title{
Time correlation functions for non-Hermitian quantum systems 
}

\author{Alessandro Sergi} 
\email{sergi@ukzn.ac.za}

\affiliation{
School of Chemistry and Physics, University of KwaZulu-Natal in Pietermaritzburg, 
Private Bag X01, Scottsville 3209, South Africa}
\affiliation{
KwaZulu-Natal Node,
National Institute for Theoretical Physics (NITheP), Durban 4000, South Africa}

\author{Konstantin G. Zloshchastiev}
\email{k.g.zloschastiev@gmail.com}

\affiliation{
Institute of Systems Science, Durban University of Technology, P.O. Box 1334, Durban 4000, South Africa}

\begin{abstract}
We introduce a formalism for time-dependent correlation
functions for systems  whose evolutions are governed
by non-Hermitian Hamiltonians of general type.
It turns out that one can define two different types
of time correlation functions.
Both these definitions seem to be physically consistent while
becoming equivalent only in certain cases.
Moreover, when autocorrelation functions are considered, 
one can introduce another function defined as the relative difference
between the two definitions.
We conjecture that such a function can be used to assess the positive
semi-definiteness of the density operator without computing
its eigenvalues.
We illustrate these points by studying analytically a number of
models with two energy levels.
\end{abstract}

\date{received: 18 December 2014 [PRA]; published: 9 June 2015 [PRA]}


\maketitle

\section{Introduction}

Quantum dynamics of systems governed by non-Hermitian (NH) Hamiltonians
is currently a very popular field of research.
For example, this includes areas such as quantum transport and scattering by 
complex potentials \cite{suu54,ck58,lay63,bender07,varga,berg,miro,varga2,muga,thila},
resonances and decaying states \cite{nimrod2,seba,spyros,fesh,sudarshan},
multiphoton ionization \cite{selsto,baker,baker2,chu}, optical waveguides \cite{optics,optics2},
and the theory of open quantum 
systems \cite{kor64,wong67,heg93,bas93,ang95,rotter,rotter2,gsz08,bellomo,banerjee,reiter,sz13,bg12,sz14}.
Notwithstanding the long history, the core of the formalism of 
non-Hermitian quantum dynamics
is still a topic of active research \cite{ghk10,ale,grsc11,sz13,bg12,sz14}.
From the viewpoint of the general formalism, one can divide the whole field of research
into two main branches: 
pseudo-Hermitian quantum dynamics (together with the synonymous or sister theories, 
such as quasi-Hermitian and ${\cal PT}$-symmetric quantum mechanics) \cite{sgh92,ben99,zno02,mos10}
and general non-Hermitian formalism.
The former branch deals mainly with a class of NH Hamiltonians that
have purely real eigenvalues.
The latter branch, to which the study reported in this paper belongs,
attempts to find a way to adopt general
NH Hamiltonians as an effective tool for the description of 
dissipative processes.
As regards this goal, the real-valuedness of the spectrum of the 
Hamiltonian operator is not compulsory - on the contrary, 
in some cases, it would be too restrictive.   
Indeed, the real parts of the eigenvalues can be related to the energy, as usual,
whereas the imaginary parts can be linked to decay rates.

In this paper, we continue along the line of research started
in Ref. \cite{sz13}. Our aim is to build
a consistent formalism of quantum statistical mechanics
when the dynamics is generated by non-Hermitian Hamiltonians.
In Ref. \cite{sz13}, we adopted a well-defined quantum evolution
equation for the density operator and considered a definition of statistical
averages that allows one to retain the probabilistic interpretation
of the density matrix.
Here, our interest is devoted to the definition and study of
the multi-time correlation functions since
they are often directly related to measurable quantities, 
such as structure factors, fluctuation spectra, and scattering rates \cite{gzbook,bpbook}.

This paper is structured as follows.
In Sec. \ref{s-nhd} we give a brief outline of the quantum non-Hermitian approach.
In Sec. \ref{s-corr-h} we sketch the formalism for
multi-time correlation functions in the conventional (Hermitian) case.
In Sec. \ref{s-corr-nh} we generalize the formalism for
multi-time correlation functions to the case of quantum dynamics
generated by non-Hermitian Hamiltonians.
In Sec. \ref{s-corr-tls} we apply such a formalism to a specific example:
we compute the two-point correlation functions for a number of 
two-level systems (TLSs), which were also considered in \cite{sz13}.
A summary is given in Sec. \ref{s-con}.

\scn{Non-Hermitian dynamics and averages}{s-nhd}

Let us consider a non-Hermitian Hamiltonian $\hat{H}\neq\hat{H}^{\dag}$,
which can be always written in the form
\begin{equation}
\hat{H}=\hat{H}_+-i\hat{\Gamma}
,
\end{equation}
where $\hat{H}_+=\hat{H}_+^{\dag}$ and $\hat{\Gamma}=\hat{\Gamma}^{\dag}$.
It can be shown that the evolution equation for the non-normalized
density matrix $\densnnorm$ can be written as
\begin{eqnarray}
\frac{d}{dt}\densnnorm(t)
&=&-\frac{i}{\hbar}\left[\hat{H}_+,\densnnorm(t)\right]
-\frac{1}{\hbar}\left\{\hat{\Gamma},\densnnorm(t)\right\}
,
\label{eq:Omega-dyna}
\end{eqnarray}
where the square brackets on the right hand side of Eq. (\ref{eq:Omega-dyna})
denote the commutator while the curly brackets denote the anti-commutator.
In the context of theory of open quantum systems the evolution equation
for the density operator $\densnnorm$
effectively describes the original 
subsystem (with Hamiltonian $\hat{H}_+$) together with the effect of environment (represented
by $\hat{\Gamma}$).
Upon taking the trace of Eq. (\ref{eq:Omega-dyna}) one obtains:
\begin{eqnarray}
\frac{d}{dt}{\rm tr} \densnnorm(t)
&=&
-\frac{2}{\hbar}{\rm tr}
\left(\densnnorm(t)\hat{\Gamma}\right)
,
\label{eq:tr-dyna}
\end{eqnarray}
which shows that the trace of $\densnnorm$ is not conserved during the evolution
governed by a NH Hamiltonian.
Defining the normalized density matrix as
\begin{equation}\lb{eq:rhonorm}
\hat\rho(t)=\frac{\densnnorm(t)}{{\rm tr} \densnnorm(t)}
,
\end{equation}
and using Eq.~(\ref{eq:tr-dyna}) we can obtain 
a non-linear equation of motion for $\hat\rho(t)$
\be
\frac{d}{dt}\hat\rho(t)
=
-\frac{i}{\hbar}\left[\hat{H}_+,\hat\rho(t)\right]
-
\frac{1}{\hbar}\left\{\hat{\Gamma},\hat\rho(t)\right\}
+\frac{2}{\hbar}\hat\rho(t){\rm tr}\left(\hat\rho(t)\hat{\Gamma}\right)
,
\label{eq:nl-ddt-rho}
\ee
which can be brought back into a linear form by using (\ref{eq:rhonorm})
as an ansatz (provided the initial condition $\text{tr} \hat\rho (t_0) = 1$ is assumed).
The evolution equation
for the normalized density operator 
effectively describes the original subsystem (with Hamiltonian $\hat{H}_+$) together with the effect of 
environment (represented
by $\hat{\Gamma}$)
and the additional term that restores the 
overall probability's conservation. 
For the detailed discussion of important features of time evolution
driven by NH Hamiltonians, 
such as the purity's non-conservation
and independence from the absolute value of the energy,
the reader is referred to Refs. \cite{sz13,sz14}. 

Given an arbitrary operator $\hat{\chi}$, its statistical average can be
defined as
\begin{equation}
\langle\hat{\chi} (t)\rangle
\equiv 
{\rm tr}\left(\hat\rho(t)\hat{\chi}\right)
=\frac{{\rm tr}\left(\densnnorm(t)\hat{\chi}\right)}
{{\rm tr}\densnnorm(t)}
,
\label{eq:ave}
\end{equation}
which reduces to the
standard quantum statistical rule in the case of Hermitian Hamiltonians.

\section{Time correlation functions}\lb{s-corr}

In the following we sketch the formalism for the two-time correlation functions
in the Hermitian case in Sec. \ref{s-corr-h}. 
Its generalization to the non-Hermitian case is given in Sec. \ref{s-corr-nh}.

\subsection{Hermitian case}\lb{s-corr-h}

Given two arbitrary operators $\hat{\chi}$ and $\hat{\xi}$, it is known that their
time-dependent correlation function in the Heisenberg picture
of Hermitian quantum mechanics is defined as
\begin{equation}
C_{\chi\xi}(t_2,t_1)\equiv\langle\hat{\chi}(t_2)\hat{\xi}(t_1)\rangle
\equiv {\rm tr}\left(\hat{\chi}(t_2)\hat{\xi}(t_1)\hat\rho(0)
\right)
.
\label{eq:herm-corf}
\end{equation}
Now, assuming that the evolution takes place under the Hermitian operator
$\hat{H}_+$, upon using the definition of the time-dependent Heisenberg operator
($\hat{\chi}(t)
\equiv\exp\left[\frac{it}{\hbar}\hat{H}_+\right]\hat{\chi}
\exp\left[\frac{-it}{\hbar}\hat{H}_+\right]$),
and the properties of the trace,
the Hermitian correlation function (\ref{eq:herm-corf})
can be rewritten as
\bw
\be
C_{\chi\xi}(t_2,t_1)
=
{\rm tr}\left\{\hat{\chi}
\exp\left[\frac{-i(t_2-t_1)}{\hbar}\hat{H}_+\right]
\hat{\xi}
\hat\rho_{\hat{H}_+}(t_1)
\exp\left[\frac{i(t_2-t_1)}{\hbar}\hat{H}_+\right]
\right\}
,
\label{eq:herm-corf2}
\ee
\ew
where
\begin{equation}
\hat\rho_{\hat{H}_+}(t_1)
\equiv 
\exp\left[\frac{-it_1}{\hbar}\hat{H}_+\right]
\hat\rho
\exp\left[\frac{it_1}{\hbar}\hat{H}_+\right]
.
\end{equation}
The definition given in Eq. (\ref{eq:herm-corf2}) represents the correlation function
written in the Schr\"odinger picture, where the time dependence has been
transferred from the operators to the density matrix.
We take the Schr\"odinger form of the correlation function as the basis
for the generalization to the non-Hermitian case, which is treated in
Sec. \ref{s-corr-nh}.

\subsection{Non-Hermitian case}\lb{s-corr-nh}

When the Hermitian Hamiltonian $\hat{H}_+$ is augmented
by a non-Hermitian part $\hat{H}_-=-i\Gamma$,
a natural generalization of Eq.~(\ref{eq:herm-corf2}) is
\ba
{\cal C}_{\xi \chi} (t_1, t_2)
&=&
 {\rm tr}\left\{\hat{\chi}
{\cal K}(t_2, t_1)
\hat{\xi} {\cal K}(t_1, t_0)\hat\rho (t_0)
\right\}
\nn\\&=&  
{\rm tr}\left\{\hat{\chi}
{\cal K}(t_2, t_1) \hat{\xi} \hat\rho (t_1) \right\}
\nn\\&= &
 {\rm tr}\left\{\hat{\chi}
{\cal K}(t_2, t_1) 
 \{
\hat{\xi} \densnnorm (t_1)/ {\rm tr} \densnnorm (t_1)
 \}
\right\}
,
\label{eq:herm-corf-nh}
\ea
where $\hat{\chi}$ and $\hat{\xi}$ are operators in the \schrod presentation,
and ${\cal K}$ is the (generalized) evolution operator defined as follows.
When ${\cal K}(t_b, t_a)$ is applied to anything on its right,
it evolves it from time $t_a$ up to time $t_b$ using Eq. (\ref{eq:nl-ddt-rho}).
Hence, in the expression above, the first application
of ${\cal K}$ evolves $\hat\rho$ from time $t_0$ to time $t_1$
as a solution of (\ref{eq:nl-ddt-rho}). 
The second application of the evolution operator 
acts on the operator $\hat{\xi} \hat\rho(t_1)$
and propagates it from the initial condition at $t_1$ until the final time
$t_2$, using Eq. (\ref{eq:nl-ddt-rho}).
Equation (\ref{eq:herm-corf-nh})
has the obvious properties that it reduces to the correlation function
of Hermitian quantum mechanics when $\hat{\Gamma}=0$ and to
the normalized average of $\hat{\chi}$ when $\hat{\xi}$ is
the identity operator.
The correlation function that is defined by  Eq. (\ref{eq:herm-corf-nh}) 
is founded on the time evolution of
the density matrix in terms of a non-Hermitian Hamiltonian
and a non-linear equation of motion.
Naturally, the non-linear equation (\ref{eq:nl-ddt-rho})
may invalidate the properties of the correlation function, which are
related to linearity.
Moreover,   
the linearizing ansatz (\ref{eq:rhonorm}), often adopted in calculations,
can be applied only if the input of the evolution operator $\cal K$
has a unit trace.
Otherwise, one should use other analytical (or numerical) approaches.

There is also the possibility of defining the correlation functions
in terms of the linear non-Hermitian evolution
given by Eq. (\ref{eq:Omega-dyna}):
\ba
{\cal C}^{(L)}_{\xi \chi} (t_1, t_2)
&=&
\frac{
 {\rm tr}\left\{\hat{\chi}
{\cal K}_L (t_2, t_1)
\hat{\xi} {\cal K}_L (t_1, t_0)\densnnorm (t_0)
\right\}
     }{
 {\rm tr}\, \densnnorm (t_2)		
     }
\nn\\&=&
\frac{
 {\rm tr}\left\{\hat{\chi}
{\cal K}_L (t_2, t_1)
\{
  \hat{\xi} \densnnorm (t_1)
\}
\right\}
     }{
 {\rm tr}\, \densnnorm (t_2)		
     }
,
\label{eq:herm-corf-nh-L}
\ea
where 
${\cal K}_L$ is the evolution operator defined as follows.
When ${\cal K}_L(t_b, t_a)$ is applied to an operator on its right,
it evolves it from time $t_a$ up to time $t_b$ using the linear equation (\ref{eq:Omega-dyna}).
Hence, in the expression above, the first application
of ${\cal K}_L$ evolves the non-normalized density matrix $\densnnorm$
from time $t_0$ to time $t_1$
as a solution of Eq. (\ref{eq:Omega-dyna}). 
The second application
of the evolution operator 
acts on the product $\hat{\xi} \densnnorm(t_1)$
and propagates it from the initial condition at $t_1$ until the final time
$t_2$, using the Eq. (\ref{eq:Omega-dyna}).
The denominator of the correlation function defined in
Eq. (\ref{eq:herm-corf-nh-L}), \emph{i.e.}, ${\rm tr}\, \densnnorm (t_2)$, 
takes into account the final
normalization since, in this case, the time evolution is realized
in terms of the non-normalized density matrix.
Also the definition in (\ref{eq:herm-corf-nh-L})
reduces to the correlation function
of Hermitian quantum mechanics when $\hat{\Gamma}=0$ and to
the normalized average of $\hat{\chi}$ when $\hat{\xi}$ is
the identity operator.
We propose Eqs. (\ref{eq:herm-corf-nh}) and (\ref{eq:herm-corf-nh-L}) as
legitimate definitions of two-time correlation functions
in the case of quantum dynamics realized by means of non-Hermitian Hamiltonians.

In the case of autocorrelation functions, when the initial time
is taken as $t_1=0$ (which is equivalent to all times being counted
from the moment $t = t_0 = t_1$),
the definitions in (\ref{eq:herm-corf-nh}) and (\ref{eq:herm-corf-nh-L}) 
can be reduced to
\ba
&&
{\cal C}_{\chi\chi} (t) 
=
 {\rm tr}\left\{\hat{\chi}
{\cal K}(t, 0) 
\hat{\chi} \hat\rho (0)
\right\}
, 
\lb{eq:acorr1-def}\\&&
{\cal C}^{(L)}_{\chi\chi} (t)
=
\frac{
 {\rm tr}\left\{\hat{\chi}
{\cal K}_L (t, 0)
  \hat{\chi} \densnnorm (0)
\right\}
     }{
 {\rm tr}\, \densnnorm (t)		
     }
,
\lb{eq:acorr2-def}
\ea
assuming that ${\rm tr}\, \densnnorm (0)= {\rm tr}\, \hat\rho (0) =1$.
We will use the definitions in Eqs. (\ref{eq:acorr1-def})
and (\ref{eq:acorr2-def}) in Sec. \ref{s-corr-tls}.

The generalization of the definitions of correlation functions in 
(\ref{eq:herm-corf-nh}) and (\ref{eq:herm-corf-nh-L})
to the multi-time case is straightforward.
In order to do this, first one can introduce the ordered sets of times 
$t_n > ... > t_1 \geqslant t_0$ and
$s_m > ... > s_1 \geqslant t_0$,
as well as their time-ordered union $\left\{\tau\right\}$:
$\tau_u > ... > \tau_1 \geqslant t_0$ where $u \leqslant n+m  $.
Next, for any set of operators 
$\hat\chi_j$ ($j = 1,..., m$) and $\hat\xi_k$ ($k = 1,..., n$) in the
\schrod picture, we define the superoperator 
$\Pi_l$ ($l=1,..., u$) through its action upon the operator $\hat D$ which can be either $\hat\rho$
or $\densnnorm$  \cite{gzbook,bpbook}:
\be
\Pi_l \hat D
= 
\left\{
\baa{lll}
\hat\xi_k \hat D & \text{if} \ \ \tau_l = t_k \not= s_j  &  \text{for some} \ k \ \text{and all} \ j ,\\
\hat D \hat\chi_j & \text{if} \ \ \tau_l = s_j \not= t_k &  \text{for some} \ j \ \text{and all} \ k ,\\
\hat\xi_k \hat D \hat\chi_j & \text{if} \ \ \tau_l = t_k = s_j  &  \text{for some} \ k \ \text{and} \ j ,
\eaa
\right.
\ee
Then in the case of the definition in (\ref{eq:herm-corf-nh}),
one can introduce the multi-time correlation functions in the standard way:
\bw
\ba
{\cal C} (t_1,..., t_n; s_1,... , s_m) 
&\equiv&
\left\langle 
\chi_1 (s_1) ... \chi_m (s_m) \xi_n (t_n) ... \xi_1 (t_1)
\right\rangle
\nn\\&=&
\text{tr}
\left\{
\Pi_u {\cal K} (\tau_u, \tau_{u-1})
\Pi_{u-1} {\cal K} (\tau_{u-1}, \tau_{u-2})
...
\Pi_{1} {\cal K} (\tau_{1}, t_{0})
\hat\rho (t_0)
\right\}
,
\label{eq:mcorf-nh}
\ea
where the evolution operator $\cal K$ is defined as in the paragraph
after Eq. (\ref{eq:herm-corf-nh}).
In the case of the other definition, given in (\ref{eq:herm-corf-nh-L}),
one can adopt the following generalization:
\ba
{\cal C}^{(L)} (t_1,..., t_n; s_1,... , s_m) 
&\equiv&
\left\langle 
\chi_1 (s_1) ... \chi_m (s_m) \xi_n (t_n) ... \xi_1 (t_1)
\right\rangle_L
\nn\\&=&
\frac{
\text{tr}
\left\{
\Pi_u {\cal K}_L (\tau_u, \tau_{u-1})
\Pi_{u-1} {\cal K}_L (\tau_{u-1}, \tau_{u-2})
...
\Pi_{1} {\cal K}_L (\tau_{1}, t_{0})
\densnnorm (t_0)
\right\}
}{
\text{tr} \densnnorm (\tau_u)
}
,
\label{eq:mcorf-nh-L}
\ea
\ew
where the evolution operator ${\cal K}_L$
is defined in the paragraph after Eq. (\ref{eq:herm-corf-nh-L}).

\begin{figure}[htbt]
\begin{center}\epsfig{figure=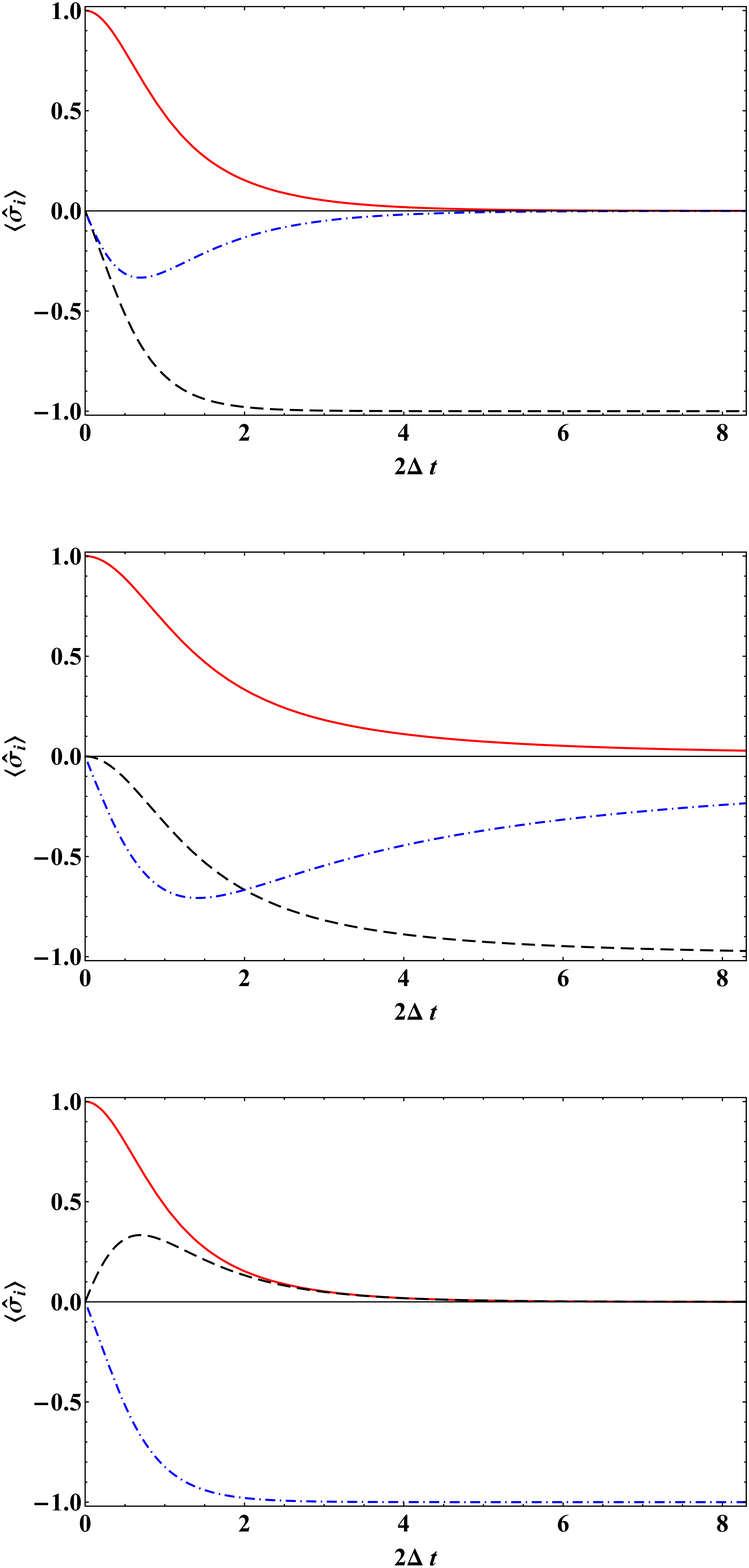,width=  0.99\columnwidth}\end{center}
\caption{(Color
online) Profiles of 
$\left\langle \hat\sigma_x \right\rangle $ (solid line),
$\left\langle \hat\sigma_y \right\rangle $ (dashed line)
and
$\left\langle \hat\sigma_z \right\rangle $ (dot-dashed line)
for the ``on-shell'' value $\nu = 0$  and, from top to bottom, $a_2 = 1, 0, -1$.  
The curves for $\acorrx$ and $\acorrx^{(L)}$ 
coincide with $\left\langle \hat\sigma_x \right\rangle $.}
\label{f-41-sigx-a2e05}
\end{figure}

Finally, it should be mentioned that
notwithstanding the mathematical equivalence of the evolution equations,
which underlie the definitions of the correlation functions,
these definitions themselves represent different physical points of view.
The definition given in Eq. (\ref{eq:mcorf-nh}) 
(whose special cases are found in Eqs. (\ref{eq:herm-corf-nh})
and (\ref{eq:acorr1-def})), 
implies that the actual time evolution is driven by the normalized
density operator $\hat\rho$ whereas the non-normalized density
$\densnnorm$ is an auxiliary quantity.
The definition given in Eq. (\ref{eq:mcorf-nh-L}) 
(whose special cases are found in Eqs. (\ref{eq:herm-corf-nh-L})
and (\ref{eq:acorr2-def})),
is based on the alternative point of view: it is the non-normalized
density operator that actually drives the time evolution.

\scn{Correlations in two-level systems}{s-corr-tls}

In this section we consider as a specific example a number of 
non-Hermitian TLS models, which were already studied in Ref. \cite{sz13}.
For these models the Hermitian part is common and is described
 by the Hamiltonian
\be
\hat H_+ = - \hbar \rabi \hat{\sigma}_x
,
\label{eq:tunnspin}
\ee
$\rabi $ being a positive-valued constant parameter,
whereas the anti-Hermitian part (the so-called decay rate operator)
$\hat\Gamma$ varies.
The models of such a kind are quite general: they are often used in order to effectively describe
the dissipative and measurement-related phenomena in open quantum-optical and spin systems,
such as the direct photodetection of a driven TLS
interacting with the electromagnetic field \cite{bpbook}.

Here
we are going to compute the correlation
functions using the definitions in Eqs. (\ref{eq:herm-corf-nh}) 
and (\ref{eq:herm-corf-nh-L}).
In particular, we are going to explicitly consider the case
of the autocorrelation functions,
defined in Eqs. (\ref{eq:acorr1-def}) and (\ref{eq:acorr2-def}).
Some other two-time correlation functions are presented in the Appendix.

With the goal of obtaining exact analytical results,
and considering the comment after Eq. (\ref{eq:herm-corf-nh}),
we restrict our study only to those initial density matrices
satisfying 
\be\lb{eq:cond-unitr}
{\rm tr} \left( \hat{\sigma}_k \hat\rho (0) \right) 
= 
{\rm tr} \left( \hat{\sigma}_k \densnnorm (0) \right)/{\rm tr} \densnnorm (0)
= 1
, 
\ee
where $k=x,z$, so that $\hat{\sigma}_k$ will denote
a Pauli matrix operator in the Schr\"odinger picture.
The initial conditions for the density matrix, which are
labelled by the index $k$, do not need to be satisfied simultaneously.
In general, such conditions restrict the physical situations that
one can describe. However, we are not aiming at providing a general
solution of the dynamics but we want to illustrate the formalism
of the correlation function, introduced in Sec. \ref{s-corr-nh},
by providing several specific examples.
Hence, for what follows,
it is useful to define the following density matrices that 
satisfy (\ref{eq:cond-unitr}) for $k = x,z$, respectively:
\ba
\hat\rho_x
&=&
\frac{1}{2}
\left(
I + \sigma_x - \nu \sigma_y
\right)
\nn\\&=&
\frac{1}{2}\sum_{j, m=g,e} |j \rangle\langle m|
+ \frac{i}{2} \nu
\left(\begin{array}{cc} 0 & ~~1 \\ -1 &~~ 0 \end{array}\right) 
,
\lb{eq:inirhox}\\
\hat\rho_z
&=&
\frac{1}{2}
\left(
I + \sigma_z - \nu \sigma_y
\right)
\nn\\&=&
|e\rangle\langle e| 
+ \frac{i}{2} \nu
\left(\begin{array}{cc} 0 & ~~1 \\ -1 &~~ 0 \end{array}\right) 
,
\lb{eq:inirhoz}
\ea
where $\nu$ is some real-valued constant parameter,
and we use the standard qubit notations 
\[
|e\rangle\langle e| =
\left(\begin{array}{cc} 1 & ~~0 \\ 0 &~~ 0 \end{array}\right) 
, \ \
\sum_{j,m=g,e} |j\rangle\langle m| =
\left(\begin{array}{cc} 1 & ~~1 \\ 1 &~~ 1 \end{array}\right) 
.
\]
In order to describe physical situations,
the initial conditions (\ref{eq:inirhox})
and (\ref{eq:inirhoz}) must be supplemented
with the condition of positive semi-definiteness 
of the density matrix $\hat\rho(0)$.
In this case, such a condition is equivalent  to
\be\lb{eq:psd}
\nu = 0
.
\ee 
The condition in Eq. (\ref{eq:psd}), relating the value of the parameter
$\nu$ to the positive semi-definiteness of the density matrix
at the initial time,
can be deduced by inspecting the eigenvalues of the density matrices
in Eqs.  (\ref{eq:inirhox}) and (\ref{eq:inirhoz}):
\be
\lambda_\pm [\hat\rho_x ]
=
\lambda_\pm [\hat\rho_z ]
=
\frac{1}{2} \left(1 \pm \sqrt{1+ \nu^2}\right)
. \label{eq:ev-do}
\ee
In what follows we will call Eq. (\ref{eq:psd}) the ``on-shell'' condition:
it makes sure that the probabilistic interpretation
of the density operator is preserved.
Hence, the parameter $\nu$ can be regarded as a measure of the deviation
of the solutions from those that are physically permitted.
However, as it will be explained in detail later, it is
convenient to first obtain the solutions ``off-shell''
(\emph{i.e.}, at $\nu \not= 0$).

\subsection{Evolution with conserved average energy (exponential decay)}\lb{s-4-1}  

Let us consider the model specified by the sum of the Hermitian
Hamiltonian in Eq. (\ref{eq:tunnspin}) and the anti-Hermitian part 
given by
\be
\hat\Gamma^{\text{(ed)}} = 
\hbar \rabi \left(a_2 \hat{\sigma}_y
+ \hat{\sigma}_z + \gamma \hat{I} \right)
,
\ee
where $a_2$ and $\gamma$ are two constant parameters.
This model was considered in section 4.1 of \cite{sz13}.

As discussed previously, 
in order to compute analytically
the autocorrelation functions $\acorrx$ and $\acorrx^{(L)}$, defined as in (\ref{eq:acorr1-def}) and (\ref{eq:acorr2-def}), 
we must impose the initial conditions in Eq. (\ref{eq:inirhox})
on the density operator.
As a result, the analytical expressions of the averages of the spin operators
have the following form:
\ba
&&
\langle \hat\sigma_x \rangle 
= 
\frac{a_2^2}{S^{(\text{ed})}_\nu (t)}
,\lb{eq:41x-sx}
\\&&
\langle \hat\sigma_y  \rangle
=\frac{1}{S^{(\text{ed})}_\nu (t)}
\biggl[ (\nu (1-a_2^2) - 1)\cosh{(\alpha t)}
\nn\\&&\qquad \qquad -a_2^2 \sinh{(\alpha t)} + 1 -\nu \biggr] ,
\lb{eq:41x-sy}
\\&&
\langle\hat\sigma_z  \rangle
=
\frac{a_2 (1-\nu)}{
S^{(\text{ed})}_\nu (t)
}  
\left( e^{- \alpha t}  - 1 
\right) 
,
\lb{eq:41x-sz}
\ea
where $\alpha = 2 a_2 \rabi$, and we introduced the function
$
S^{(\text{ed})}_b (t)
=
(a_2^2 - b  +1) \cosh{(\alpha t)} +  a_2^2 b \sinh{(\alpha t)}  + b -1
,
$
with $b$ being an index.

It is worth noting that the observable values do not depend on the parameter 
$\gamma$.
This arises from the independence of Eq. (\ref{eq:nl-ddt-rho})
on the absolute value of the energy,
as it was already discussed in \cite{sz13,sz14}.
Another thing to note is that the value $\langle \hat\sigma_x \rangle
\propto \left\langle H_+\right\rangle$,
though vanishing at long times (as shown below),
is not identically zero, since the initial condition (\ref{eq:inirhox}) 
was used. Such a condition is different
from the one used in \cite{sz13}.
The form of the autocorrelation functions is:
\ba
&&
\acorrx (t)
=
a_2^2
\biggl[
         \left(a_2^2 + i \nu a_2  +1 \right) \cosh{(\alpha t)} 
\nn\\&&\qquad \qquad \quad + i \nu a_2 \sinh{(\alpha t)} - i \nu a_2 -1
\biggr]^{-1} ,
\lb{eq:41x-ac}
\\&&
\acorrx^{(L)} (t)
=
\frac{a_2^2}{S^{(\text{ed})}_\nu (t)}
=
\langle \hat\sigma_x \rangle 
.
\lb{eq:41x-acl}
\ea
The long-time asymptotic values of 
Eqs. (\ref{eq:41x-sx}) - (\ref{eq:41x-acl})
are:
\ba
&&
\lim\limits_{t\to +\infty}\langle \hat\sigma_x \rangle
=
0
,\\&&
\lim\limits_{t\to +\infty}
\langle \hat\sigma_y \rangle
= 
-
\left(
\frac{1-a_2^2}{1+a_2^2}
\right)^{\theta (-\alpha)}
,\\&&
\lim\limits_{t\to +\infty}
\langle \hat\sigma_z \rangle
= 
\frac{2 \theta (-\alpha) a_2}{1+a_2^2}
,\\&&
\lim\limits_{t\to +\infty}
\acorrx (t)
=
\lim\limits_{t\to +\infty}
\acorrx^{(L)} (t)
=
0
,
\ea
where $\theta (x)$ is the Heaviside step function.
The comparative plots of different observables for some values
of the parameters are shown in Fig. \ref{f-41-sigx-a2e05}.

In order to compute analytically the autocorrelation functions
$\acorrz$ and $\acorrz^{(L)}$, the initial condition
in Eq. (\ref{eq:inirhoz}) is imposed.
As a result, the analytical expressions of the averages of the spin operators
have the form:
\ba
&&
\langle \hat\sigma_x \rangle = 0 
,
\lb{eq:41z-sx}
\\&&
\langle \hat\sigma_y  \rangle
=\frac{a_2 -1}{T^{(\text{ed})}_\nu (t)}
\biggl[ (1 - \nu (a_2 + 1))\cosh{(\alpha t)} 
\nn\\&& \qquad \quad  -a_2\sinh{(\alpha t)} - \frac{\nu}{a_2 -1} -1\biggr] ,
\\&&
\langle\hat\sigma_z  \rangle
=
\frac{a_2 (1-\nu)}{
T^{(\text{ed})}_\nu (t)
}  
\left( e^{- \alpha t} + \frac{a_2}{1-\nu} - 1 
\right) 
,
\ea
where we introduced the function
$
T^{(\text{ed})}_b (t)
=
(a_2^2 - a_2 - b  +1) \cosh{(\alpha t)} - a_2 (1 - a_2 b)
\sinh{(\alpha t)} +a_2  + b -1
.
$
The form of the corresponding autocorrelation functions is:
\ba
&&
\acorrz (t)
=\frac{a_2}{T^{(\text{ed})}_0 (t)}
\left(
         e^{- \alpha t} + a_2 - 1
\right) ,
\\&&
\acorrz^{(L)} (t)
=\frac{a_2}{T^{(\text{ed})}_\nu (t)}
\left(
         e^{- \alpha t} + a_2 - 1
\right)
.
\lb{eq:41z-acl}
\ea
The long-times asymptotic values of 
Eqs. (\ref{eq:41z-sx}) - (\ref{eq:41z-acl})
are:
\ba
&&
\lim\limits_{t\to +\infty}\langle \hat\sigma_x \rangle
=
0
,\\&&
\lim\limits_{t\to +\infty}
\langle \hat\sigma_y \rangle
= 
-
\left(
\frac{1-a_2^2}{1+a_2^2}
\right)^{\theta (-\alpha)}
,\\&&
\lim\limits_{t\to +\infty}
\langle \hat\sigma_z \rangle
= 
\frac{2 \theta (-\alpha) a_2}{1+a_2^2}
,\\&&
\lim\limits_{t\to +\infty}
\acorrz (t)
=
\frac{2 \theta (-\alpha) a_2}{1+a_2^2}
,\\&&
\lim\limits_{t\to +\infty}
\acorrz^{(L)} (t)
=
\frac{2 \theta (-\alpha) a_2}{1+a_2^2}
\frac{1}{1-\nu}
.
\ea
In this set of solutions, it is clear that the autocorrelation
function $\acorrz^{(L)} (t)$
depends on the initial conditions.
It is also worth noticing that the ratio
\[ 
\frac{
\lim\limits_{t\to +\infty} \acorrz (t) 
}{
\lim\limits_{t\to +\infty} \acorrz^{(L)} (t) 
}
= 1-\nu
\]
does not depend on the parameters of the Hamiltonian but
it depends on the parameter $\nu$ whose nature was discussed after
Eq. (\ref{eq:ev-do}).

The comparative plots of different observables for some values
of parameters are shown in Fig. \ref{f-41-sigz-a2e05}.
It is easy to see that the difference between the two functions
${\cal C}_{\sigma_k\sigma_k} (t)$, 
and ${\cal C}^{(L)}_{\sigma_k\sigma_k} (t)$, 
for each case $k=x,z$, disappears when $\nu \to 0$.
The implications of this feature are discussed in Sec. \ref{s-con}.

\subsection{Evolution with conserved average energy (polynomial decay)}\lb{s-4-2}  

Here, we consider the model specified by the sum of the Hermitian
Hamiltonian Eq. (\ref{eq:tunnspin})
and the anti-Hermitian part given by
\be
\hat\Gamma^{\text{(pd)}} = 
\hbar \rabi 
\left(
\hat{\sigma}_z + \gamma \hat{I} \right)
,
\ee
where the parameter $\gamma$ is specified as in Sec. \ref{s-4-1}.
This model was considered in section 4.2 of \cite{sz13}.

\begin{figure}[htbt]
\begin{center}\epsfig{figure=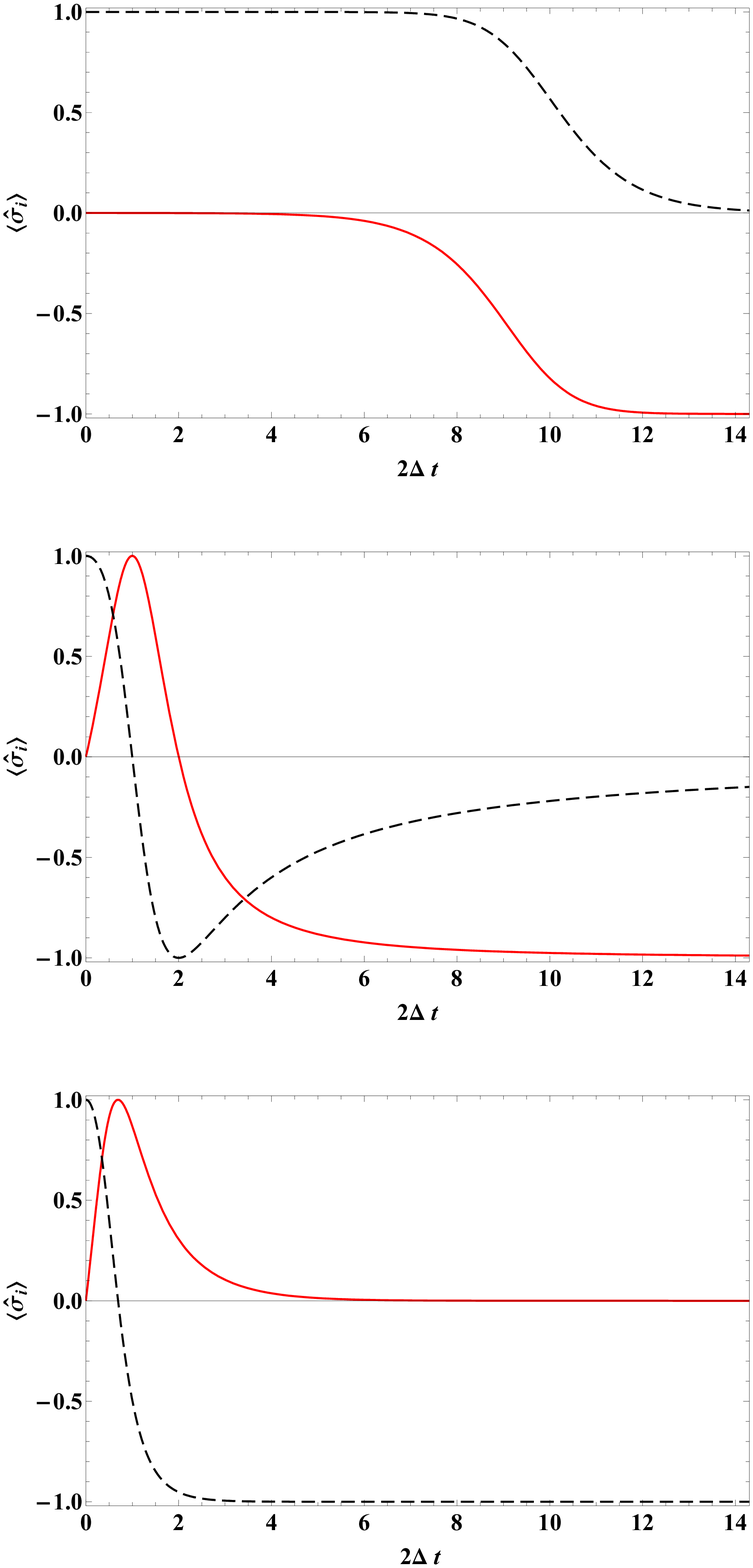,width=  0.99\columnwidth}\end{center}
\caption{(Color
online) Profiles of 
$\left\langle \hat\sigma_y \right\rangle $ (solid line)
and
$\left\langle \hat\sigma_z \right\rangle $ (dashed line)
for the ``on-shell'' value $\nu = 0$  and, from top to bottom, $a_2 = 1.0001, 0, -1$.  
The value  $\left\langle \hat\sigma_x \right\rangle$ 
is identically zero, and the curves for $\acorrx$ and $\acorrx^{(L)}$ 
coincide with $\left\langle \hat\sigma_z \right\rangle $.}
\label{f-41-sigz-a2e05}
\end{figure}

As discussed previously, 
in order to compute analytically
the autocorrelation functions $\acorrx$ and $\acorrx^{(L)}$,
we use  the initial condition specified by Eq. (\ref{eq:inirhox})
on the density operator.
Consequently,
the analytical expressions of the averages of the spin operators
have the form
\ba
&&
\langle \hat\sigma_x \rangle = 0 
,
\lb{eq:42x-sx}
\\&&
\langle \hat\sigma_y  \rangle
=
\frac{1 - \nu}{ S^{(\text{pd})}_\nu (t) }
-1 
,
\\&&
\langle\hat\sigma_z  \rangle
=
\frac{2 \rabi (\nu -1)  t}{ S^{(\text{pd})}_\nu (t) } 
,
\ea
where we introduced the function
$
S^{(\text{pd})}_b (t)
=
2 \rabi^2 (1- b)  t^2 +1
$.
The form of the autocorrelation functions is:
\ba
&&
\acorrx (t)
=
\left[
S^{(\text{pd})}_0 (t)
+ 
2 i \nu \rabi \, t
\right]^{-1}
,
\\&&
\acorrx^{(L)} (t)
=
1/S^{(\text{pd})}_\nu (t)
.
\lb{eq:42x-acl}
\ea
The long-times asymptotic values of 
Eqs. (\ref{eq:42x-sx}) - (\ref{eq:42x-acl})
are:
\ba
&&
\lim\limits_{t\to +\infty}\langle \hat\sigma_x \rangle
=
\lim\limits_{t\to +\infty}
\langle \hat\sigma_z \rangle
= 
0
,\\&&
\lim\limits_{t\to +\infty}
\langle \hat\sigma_y \rangle
= 
- 1
,\\&&
\lim\limits_{t\to +\infty}
\acorrx (t)
=
\lim\limits_{t\to +\infty}
\acorrx^{(L)} (t)
=
0
.
\ea
The graphs of different observables, corresponding to some values
of the parameters, can be found in the middle plot of
Fig. \ref{f-41-sigx-a2e05}.

In order to compute analytically the autocorrelation functions
$\acorrz$ and $\acorrz^{(L)}$,
we impose the initial condition given by Eq. (\ref{eq:inirhoz}) 
on the density operator.
It follows that the analytical expressions of the averages of 
the spin operators are:
\ba
&&
\langle \hat\sigma_x \rangle = 0 
,
\lb{eq:42z-sx}
\\&&
\langle \hat\sigma_y  \rangle
=
\frac{1 - \nu}{ T^{(\text{pd})}_\nu (t) }
-1 
,
\\&&
\langle\hat\sigma_z  \rangle
=
\frac{1-2 \rabi (1-\nu)  t}{ T^{(\text{pd})}_\nu (t) } 
,
\ea
where 
$
T^{(\text{pd})}_b (t)
=  
2\rabi  
\left[ \rabi (1 - b)  t -1 \right] t +1 
$.
The autocorrelation functions have the form
\ba
&&
\acorrz (t)
=
\frac{1 - 2 \rabi \, t}{T^{(\text{pd})}_0 (t)}
,
\\&&
\acorrz^{(L)} (t)
=
\frac{1 - 2 \rabi \, t}{T^{(\text{pd})}_\nu (t)}
.
\lb{eq:42z-acl}
\ea

The long-times asymptotic values of Eqs. (\ref{eq:42z-sx}) - (\ref{eq:42z-acl}) are:
\ba
&&
\lim\limits_{t\to +\infty}\langle \hat\sigma_x \rangle
=
\lim\limits_{t\to +\infty}
\langle \hat\sigma_z \rangle
= 
0
,\\&&
\lim\limits_{t\to +\infty}
\langle \hat\sigma_y \rangle
= 
- 1
,\\&&
\lim\limits_{t\to +\infty}
\acorrz (t)
=
0
,\\&&
\lim\limits_{t\to +\infty}
\acorrz^{(L)} (t)
=
\delta_{1 \nu}
,
\ea
where $\delta_{a b}$ is the Kronecker symbol.

The graphs of different observables, corresponding to some values
of the parameters, can be found in the middle plot in
Fig. \ref{f-41-sigz-a2e05}.
As before, the difference between the two functions ${\cal C}_{\sigma_k \sigma_k} (t)$
and ${\cal C}^{(L)}_{\sigma_k \sigma_k} (t)$ for each case $k = x, z$ disappears when $\nu \to 0$.
The implications of this feature are discussed in Sec. \ref{s-con}.

\begin{figure}[htbt]
\begin{center}\epsfig{figure=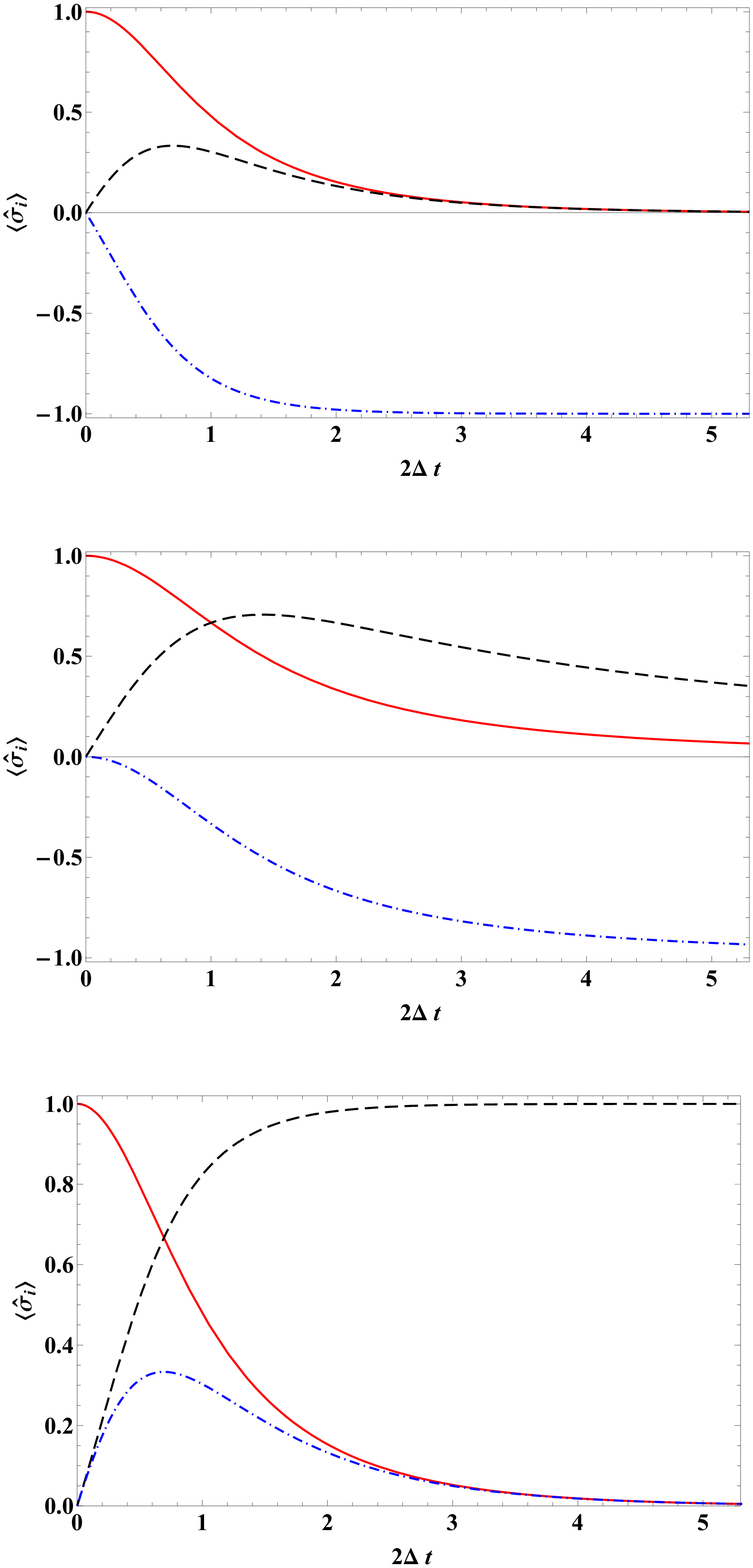,width=  0.99\columnwidth}\end{center}
\caption{(Color
online) Profiles of 
$\left\langle \hat\sigma_x \right\rangle $ (solid line),
$\left\langle \hat\sigma_y \right\rangle $ (dashed line)
and
$\left\langle \hat\sigma_z \right\rangle $ (dot-dashed line)
for the ``on-shell'' value $\nu = 0$  and, from top to bottom, $\gamma = 1, 0, -1$.  
The curves for $\acorrx$ and $\acorrx^{(L)}$ 
coincide with $\left\langle \hat\sigma_x \right\rangle $.}
\label{f-43-sigx-gammae05}
\end{figure}

\subsection{Evolution with asymptotic dephasing}\lb{s-6}  

Let us consider the model specified by the sum of the Hermitian
Hamiltonian in Eq. (\ref{eq:tunnspin}) and the anti-Hermitian part 
given by
\be
\hat\Gamma^{(\text{dph})}
=
-
\hbar  \rabi
\left[
\hat{\sigma}_y
-
\gamma
\left(
\hat{\sigma}_z 
+
\hat{I}
\right)
\right]
, \label{eq:mod-dph}
\ee
where $\gamma $ is a constant parameter. 
This model was proposed in section 6 of \cite{sz13}.
Since in Eq. (\ref{eq:mod-dph}) the parameter $\gamma $ does not appear only  
in the $\hat I$ term,
it is going to appear in all of the expressions for the observables.

As discussed previously, 
in order to compute analytically
the autocorrelation functions $\acorrx$ and $\acorrx^{(L)}$,
we use  the initial condition specified by Eq. (\ref{eq:inirhox})
on the density operator.  Consequently, solving the evolution equation yields 
the following components of the (non-normalized) density matrix:
\ba
&&
(\densnnorm)_{11} = \frac{1}{2} e^{- 2 \Gamma t} \;,
\\&&
(\densnnorm)_{12}
=(\densnnorm)_{21}^* = 
\frac{e^{- 2 \Gamma t}}{2 \gamma} 
\left[ (\gamma + i (\nu\gamma -1)) e^{\Gamma t} + i 
\right]\!
,~~~
\\&&
(\densnnorm)_{22}
=
\frac{
\left(e^{-\Gamma t}-1\right)^2
+
2 \nu \gamma \left(e^{-\Gamma t}-1 \right)
+\gamma^2
}{2\gamma^2}
,
\ea
where $\Gamma = 2 \gamma \rabi$.
It is worth recalling that for the
model given by Eqs. (\ref{eq:tunnspin}) and (\ref{eq:mod-dph}),
the off-diagonal components of the density matrix
vanish at long times.
Furthermore, the analytical expressions of the averages of the spin operators
have the form:
\ba
&&
\langle \hat\sigma_x \rangle
=
\frac{
\gamma^2
}{
S^{(\text{dph})}_\nu (t)
}
,
\lb{eq:6x-sx}
\\&&
\langle \hat\sigma_y \rangle
=
\frac{
\gamma
}{
S^{(\text{dph})}_\nu (t)
}
\left( 1 - \nu \gamma - e^{-\Gamma t}
\right)
,\\&&
\langle \hat\sigma_z \rangle
=
\frac{
\gamma^2 e^{-\Gamma t}
}{
S^{(\text{dph})}_\nu (t)
}
-
1
,
\ea 
where $\tilde\gamma^2 = \gamma^2  +1$ and
we have introduced the function
$
S^{(\text{dph})}_b (t)
=
(\tilde\gamma^2 - b \gamma) \cosh{(\Gamma t)} 
- b \gamma \sinh{(\Gamma t)} 
+ b \gamma -1
.
$
The autocorrelation functions have the form:
\be
\acorrx (t)
=
\frac{\gamma^2
}{
S^{(\text{dph})}_0 (t) 
+ 
i \nu 
\left[
\gamma^2 \sinh{(\Gamma t)}
-
\cosh{(\Gamma t)} 
+1
\right]
}
,
\ee
\be
\acorrx^{(L)} (t)
=
\frac{\gamma^2}{S^{(\text{dph})}_\nu (t)}
.
\lb{eq:6x-acl}
\ee
The long-time asymptotic values of Eqs. (\ref{eq:6x-sx}-\ref{eq:6x-acl}) are:
\ba
&&
\lim\limits_{t\to +\infty}\langle \hat\sigma_x \rangle
=
0
,\\&&
\lim\limits_{t\to +\infty}
\langle \hat\sigma_y \rangle
= 
-
\frac{2 \gamma \theta (- \Gamma) }{\tilde\gamma^2}
,\\&&
\lim\limits_{t\to +\infty}
\langle \hat\sigma_z \rangle
= 
-
\left(
\frac{1 - \gamma^2 }{\tilde\gamma^2}
\right)^{\theta (- \Gamma)}\!,~~~ \\&&
\lim\limits_{t\to +\infty}
\acorrx (t)
=
\lim\limits_{t\to +\infty}
\acorrx^{(L)} (t)
=
0
,
\ea
where the function $\theta (x)$ has been defined in Sec. \ref{s-4-1}.
The comparative plots of different observables for some values
of parameters are shown in Fig. \ref{f-43-sigx-gammae05}.

In order to compute analytically the autocorrelation functions
$\acorrz$ and $\acorrz^{(L)}$, the initial condition
in Eq. (\ref{eq:inirhoz}) is imposed.
As a result, the analytical expressions of the averages of the spin operators
have the form:
\ba
&&
\langle \hat\sigma_x \rangle
=
0
,
\lb{eq:6z-sx}
\\&&
\langle \hat\sigma_y \rangle
=
\frac{
\gamma
}{
T^{(\text{dph})}_\nu (t)
}
\left( 2 - \nu \gamma - 2 e^{-\Gamma t}
\right)
,\\&&
\langle \hat\sigma_z \rangle
=
\frac{
2 \gamma^2 e^{-\Gamma t}
}{
T^{(\text{dph})}_\nu (t)
}
-
1
,
\ea 
where 
we have introduced the function
$
T^{(\text{dph})}_b (t)
=
(\tilde\gamma^2 - b \gamma + 1) \cosh{(\Gamma t)} 
- \gamma (\gamma + b) \sinh{(\Gamma t)} 
+ b \gamma -2
.
$
The autocorrelation functions have the form:
\ba
&&
\acorrz (t)
=
\frac{4(1- \cosh{(\Gamma t)})}{
T^{(\text{dph})}_0 (t) 
}
+1
,
\\&&
\acorrz^{(L)} (t)
=
\frac{T^{(\text{dph})}_0 + 4(1- \cosh{(\Gamma t)})}{T^{(\text{dph})}_\nu (t)}
.
\lb{eq:6z-acl}
\ea
The long-time asymptotic values of Eqs. (\ref{eq:6z-sx}-\ref{eq:6z-acl}) are:
\ba
&&
\lim\limits_{t\to +\infty}\langle \hat\sigma_x \rangle
=
0
,\\&&
\lim\limits_{t\to +\infty}
\langle \hat\sigma_y \rangle
= 
-
\frac{2 \gamma \theta (- \Gamma) }{\tilde\gamma^2}
,\\&&
\lim\limits_{t\to +\infty}
\langle \hat\sigma_z \rangle
= 
\lim\limits_{t\to +\infty}
\acorrz (t)
=
-
\left(
\frac{1 - \gamma^2 }{\tilde\gamma^2}
\right)^{\theta (- \Gamma)}\!,~~~~~\\&&
\lim\limits_{t\to +\infty}
\acorrz^{(L)} (t)
=
\begin{cases}
(\nu \gamma - 1)^{-1} & \text{if } \Gamma > 0 , \\
\left(
1 - \gamma^2 
\right)/ \tilde\gamma^2 & \text{if } \Gamma < 0 ,
\end{cases}
\ea
where $\theta (x)$ has been defined in Sec. \ref{s-4-1}. 

\begin{figure}[htbt]
\begin{center}\epsfig{figure=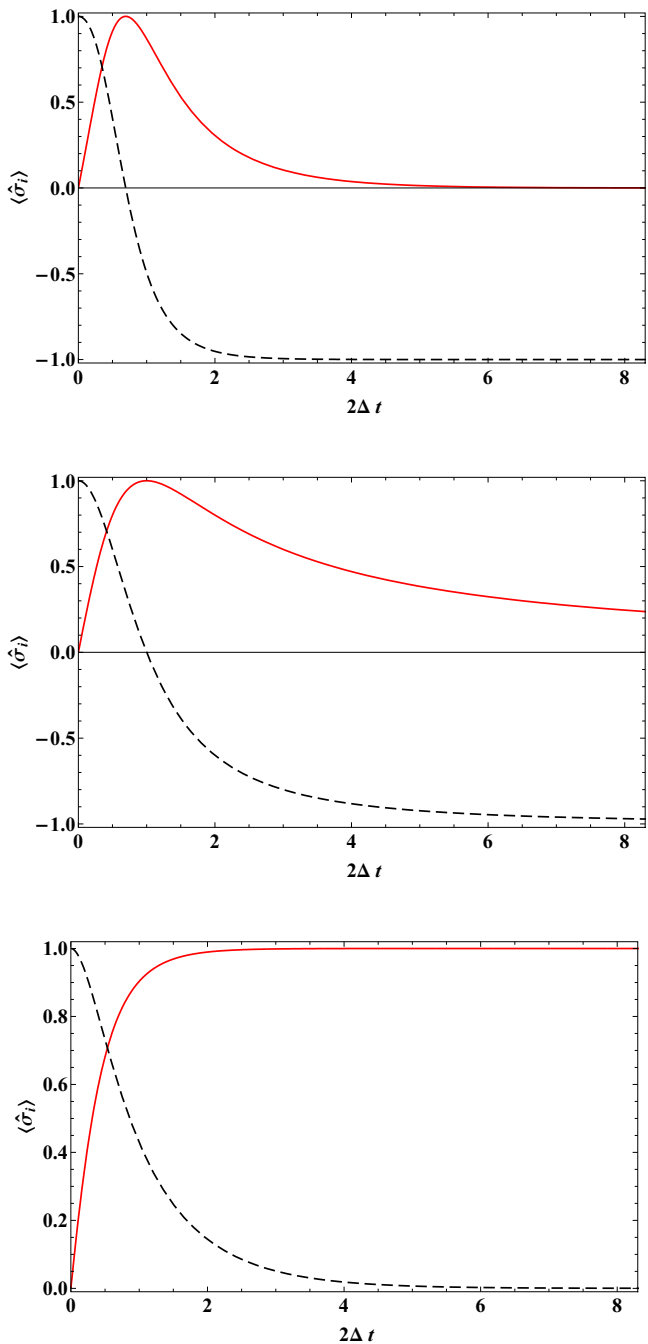,width=  0.99\columnwidth}\end{center}
\caption{(Color
online) Profiles of 
$\left\langle \hat\sigma_y \right\rangle $ (solid line)
and
$\left\langle \hat\sigma_z \right\rangle $ (dashed line)
for the ``on-shell'' value $\nu = 0$  and, from top to bottom, $\gamma = 1, 0, -1$.  
The value  $\left\langle \hat\sigma_x \right\rangle$ 
is identically zero, and the curves for $\acorrx$ and $\acorrx^{(L)}$ 
coincide with $\left\langle \hat\sigma_z \right\rangle $.}
\label{f-43-sigz-gammae05}
\end{figure}

The comparative plots of different observables for some values
of parameters are shown in Fig. \ref{f-43-sigz-gammae05}.
As in the previous sections, the difference between the two functions  ${\cal C}_{\sigma_k \sigma_k} (t)$
and ${\cal C}^{(L)}_{\sigma_k \sigma_k} (t)$ for each case $k = x, z$ disappears when $\nu \to 0$.
The possible implications of this feature shall be discussed in the next section.

\scn{Discussion and Conclusion}{s-con}

Having as a final goal the proper foundation of the statistical mechanics
of systems with non-Hermitian Hamiltonians (which is a research endeavor 
that we started in Ref. \cite{sz13}), in this paper 
we have introduced a formalism for multi-time correlation functions.
The approach is general: it depends neither on the number of degrees of freedom in the system nor on the dimensionality of the Hilbert space itself; it is also valid if the degrees of freedom are continuous.
We have found that such a formalism can lead 
to two different definitions of the time correlation function.
Notwithstanding the equivalence of the evolution equations underlying
these definitions, the different time correlation functions 
represent distinct physical points of view.
The first definition is presented in Eq. (\ref{eq:mcorf-nh})
and its special cases are given
by Eqs. (\ref{eq:herm-corf-nh}) and (\ref{eq:acorr1-def}).
Such a definition
presumes that the actual dynamics is defined in terms of the normalized
density operator.
The second definition is presented in Eq. (\ref{eq:mcorf-nh-L})
and its special cases are given 
by Eqs. (\ref{eq:herm-corf-nh-L}) and (\ref{eq:acorr2-def}).
Alternatively, this second definition assumes that the dynamics
is defined in terms of the non-normalized density operator.

In the case of one-point time correlation functions (which provide basically
statistical averages), both definitions
lead to the same results.
However, in general (\emph{i.e.}, for multiple-time correlation functions),
the two definitions do not coincide, as it has been illustrated
by explicitly studying various two-level models.

As we have already remarked at the end of Secs. \ref{s-4-1}-\ref{s-6},
the relative difference between the two definitions of autocorrelation
functions, 
$
\Delta {\cal C}_i 
= 
1-
{\cal C}_{\sigma_i \sigma_i}  / {\cal C}_{\sigma_i \sigma_i}^{(L)} 
$
($i=x,y,z$),
tends to zero when $\nu$ approaches the ``on-shell'' value (\ref{eq:psd}), at any time $t$.
As it is known, the probabilistic interpretation of a density operator
requires the matrices (\ref{eq:inirhox}) and (\ref{eq:inirhoz})
to be positive semi-definite.
The parameter $\nu$ indicates the deviation from such a property,
as it has been discussed after Eq. (\ref{eq:ev-do}).
Therefore, one can naturally propose the conjecture that
functions such as $\Delta {\cal C}_i $ are useful in order to assess
the positive semi-definiteness of the density operator
without actually computing its eigenvalues.
In particular, such a conjecture could be especially useful when the number of density matrix's
eigenvalues is rather large (\emph{e.g.}, when the Hilbert space is infinite).
In this paper, we have verified by studying various two-level models
that, indeed, the positive semi-definiteness of the density matrix
holds whenever the functions $\Delta {\cal C}_i $ vanish.
The further investigation of this problem is 
an interesting direction of future work.

\section*{Acknowledgments}

Fruitful discussions with
Ilya Sinayskiy, Hermann Uys, Mauritz van Den Worm, Vitalii Semin 
and other participants of the conference
``Quantum Information Processing, Communication and Control 3''
(3-7 November, 2014, KwaZulu-Natal, South Africa),
where parts of this work have been presented, are gratefully acknowledged.
This research was supported by
the National Research Foundation of South Africa.

\appendix*
\scn{Other correlation functions for TLS}{app} 

Apart from the autocorrelation functions computed in Sec. \ref{s-corr-tls},
one can of course compute  other types of correlation functions,
for each of the two-level systems we have considered above.
For instance, below we present the two-time correlations $\corrxz$, $\corrxz^{(L)}$, $\corryz$ and $\corryz^{(L)}$,
where we
assume the notation:
\ba
&&
{\cal C}_{\xi\chi} (t) 
=
 {\rm tr}\left\{\hat{\chi}
{\cal K}(t, 0) 
\hat{\xi} \hat\rho (0)
\right\}
, 
\lb{eq:corr1-def}\\&&
{\cal C}^{(L)}_{\xi\chi} (t)
=
\frac{
 {\rm tr}\left\{\hat{\chi}
{\cal K}_L (t, 0)
  \hat{\xi} \densnnorm (0)
\right\}
     }{
 {\rm tr}\, \densnnorm (t)		
     }
,
\lb{eq:corr2-def}
\ea
and set the initial time to zero, which is equivalent to all times being counted
from the moment $t = t_0 = t_1$.

Further, as long as the operator, which is first from the right in the definitions of functions
$\corrxz$, $\corrxz^{(L)}$, $\corryz$ and $\corryz^{(L)}$, is 
$\hat\sigma_z$, below we are going to use $\hat\rho_z$, as defined in Eq. (\ref{eq:inirhoz}),
as the initial value
for the density operator, for the reasons specified above.

\sscn{Evolution with conserved average energy (exponential decay)}{app-4-1} 

For the model from Sec. \ref{s-4-1} we obtain
\ba
&&
\corrxz (t)
=
\frac{i a_2^2 \nu}{T^{(\text{ed})}_0 (t)}
,
\\&&
\corrxz^{(L)} (t)
=\frac{i a_2^2 \nu}{T^{(\text{ed})}_\nu (t)}
,
\ea
and
\ba
&&
\corryz (t)
=
\frac{a_2 -1}{T^{(\text{ed})}_0 (t)}
\left[ \cosh{(\alpha t)}-a_2\sinh{(\alpha t)} -1 \right]\!,~~~~~~
\\&&
\corryz^{(L)} (t)
=\frac{a_2 - 1}{T^{(\text{ed})}_\nu (t)}
\left[ \cosh{(\alpha t)}-a_2\sinh{(\alpha t)} -1 \right]\!,
\ea
where we have used the notation from Sec. \ref{s-4-1}.

\sscn{Evolution with conserved average energy (polynomial decay)}{app-4-2} 

For the model from Sec. \ref{s-4-2} we obtain
\ba
&&
\corrxz (t)
=
\frac{i \nu}{T^{(\text{pd})}_0 (t)}
,
\\&&
\corrxz^{(L)} (t)
=\frac{i \nu}{T^{(\text{pd})}_\nu (t)}
,
\ea
and
\ba
&&
\corryz (t)
=
\frac{2 \rabi (1 - \rabi \, t) t}{T^{(\text{pd})}_0 (t)}
,
\\&&
\corryz^{(L)} (t)
=
\frac{2 \rabi (1 - \rabi \, t) t}{T^{(\text{pd})}_\nu (t)}
,
\ea
where the notation of Sec. \ref{s-4-2} is implied.

\sscn{Evolution with asymptotic dephasing}{app-6}

For the model from Sec. \ref{s-6} we obtain
\ba
&&
\corrxz (t)
=
\frac{i \gamma^2 \nu}{T^{(\text{dph})}_0 (t)}
,
\\&&
\corrxz^{(L)} (t)
=
\frac{i \gamma^2 \nu}{T^{(\text{dph})}_\nu (t)}
,
\ea
and
\ba
&&
\corryz (t)
=
\frac{2 \gamma (1 - e^{-\Gamma t})}{T^{(\text{dph})}_0 (t)}
,
\\&&
\corryz^{(L)} (t)
=
\frac{2 \gamma (1 - e^{-\Gamma t})}{T^{(\text{dph})}_\nu (t)}
,
\ea
where the notation of Sec. \ref{s-6} is implied.


\end{document}